\documentclass[12pt]{article}

\usepackage{amsmath}
\usepackage{amssymb}

\usepackage{graphicx}

\usepackage{setspace} 
\usepackage[labelfont=bf,labelsep=period,justification=raggedright]{caption}

\usepackage{cell}
\date{}

\begin{document}

\begin{flushleft}
{\Large
\textbf{Genetic redundancies enhance information transfer in noisy regulatory circuits}
}

{\large
Guillermo Rodrigo$^{1}$ and Juan F. Poyatos$^2\dagger$}\\


$^1$ Instituto de Biolog\'ia Molecular y Celular de Plantas, CSIC -- UPV, 46022 Valencia, Spain, $^2$ Logic of Genomic Systems Laboratory (CNB-CSIC), 28049 Madrid, Spain. \\

\vspace{.8cm}
$^\dagger$ Corresponding author. Logic of Genomic Systems Laboratory (CNB-CSIC), Darwin 3, Campus de Cantoblanco, 28049 Madrid, Spain. \\
E-mail: jpoyatos@cnb.csic.es. 

\end{flushleft}

\newpage
\section*{Abstract}
Cellular decision making is based on regulatory circuits that associate signal thresholds to specific physiological actions. This transmission of information is subjected to molecular noise what can decrease its fidelity. Here, we show instead how such intrinsic noise enhances information transfer in the presence of multiple circuit copies. The result is due to the contribution of noise to the generation of autonomous responses by each copy, which are altogether associated with a common decision. Moreover, factors that correlate the responses of the redundant units (extrinsic noise or regulatory cross-talk) contribute to reduce fidelity, while those that further uncouple them (heterogeneity within the copies) can lead to stronger information gain. Overall, our study emphasizes how the interplay of signal thresholding, redundancy, and noise influences the accuracy of cellular decision making. Understanding this interplay provides a basis to explain collective cell signaling mechanisms, and to engineer robust decisions with noisy genetic circuits.

\newpage
\section*{INTRODUCTION}
\noindent

The biochemistry of cells determines the operation of biological circuits. This biochemistry is inevitable noisy~\cite{mcadams1997,elowitz2002,lestas2010} what immediately suggests a limitation to the reliable function of these circuits, and thus many early studies examined how the problem of achieving correct operation could nevertheless be solved. Mechanisms such as kinetic proofreading~\cite{hopfield1974}, or integral feedback control~\cite{yi2000} emerged then as some fundamental solutions. One might ask, on the other hand, to what extent noise could indirectly represent an advantage. An example is found when cell populations, in which noise leads to phenotypic variability, display heterogeneity in stress responses that represent a crucial element for survival, e.g.,~\cite{bishop2007}.

In a more direct situation, noise can turn into a indispensable ingredient to facilitate new classes of behaviors not achievable otherwise~\cite{balaban2004,suel2006,acar2008,turcotte2008,raj2010}. These valuable behaviors are typically related to cellular decisions, which essentially involve changes in the expression phenotype. Specific biological circuits were consequently shown to employ noise to induce the expression of transient phenotypes~\cite{suel2006}, or to switch among distinct stable states~\cite{acar2008}. That many of these probabilistic dynamics relate to systems whose actions are susceptible to limiting signal values~\cite{feinerman2008} emphasizes the connection between noise, cellular decisions, and threshold response circuits.

The beneficial aspect of noise also forces us to revisit some of the early arguments on the relationship between stochasticity and the structure of biological systems~\cite{lerner1954,mcadams1999}. In particular, the existence of genetic redundancies was typically interpreted as a mean to enhance reliability of operation (i.e., noise as a disruptive element). This role appeared in consequence as a plausible rationale for the evolutionary maintenance of several copies of a gene or circuit~\cite{nowak1997}. Instead, we focus here on redundancy as a genetic architecture that, when coupled to the effect of noise in threshold response circuits, enables unique information-processing functions.

We examined this issue within the precise framework of information theory. Biological circuits are in this way interpreted as communication channels, in which an input signal ($x$) originates --as a result of a cellular decision-- an expression output ($y$), with a given probability (Fig.~1A). The uncertainty on the input signal is then reduced by the decision process, whose set of outcomes tells us about the input distribution~\cite{levchenko2014,bowsher2014}. This association is properly quantified by the mutual information (MI), an information-theoretic measure describing the correlation between the input signal and the output phenotype (Fig.~1A). Notably, this framework was recently exploited to quantify the functionality of transcriptional regulatory elements~\cite{tkacik2008,libby2007,richard2008}, the accuracy of cell location during developmental processes~\cite{dubuis2013}, and the maximal information transmission capacity of noisy signaling pathways~\cite{cheong2011,hansen2015}. The relevance of redundancies was already manifested in some of these results.

Here, we first illustrate how intrinsic noise (from stochastic biochemical reactions) can help to gain information. We then show that information transfer can be amplified, if the combined response of multiple genetic units is considered. The reported amplification is shown to rely on the presence of different factors that contribute to generate variability in the individual response of each unit, like intrinsic noise or genetic heterogeneity (i.e., differences in the biochemical properties). This variability helps to enlarge the capacity of the global output to represent the input distribution. In contrast, we also discuss how factors reducing variability in the responses, like a noise source common to all units (extrinsic noise) or regulatory cross-talk, eventually mitigate the gain.

\section*{RESULTS}

\subsection*{Intrinsic noise can amplify information transfer} 

We first analyzed a minimal regulatory circuit implemented by a gene (whose expression we denote as $y$) autoactivating transcriptionally its own production~\cite{wall2004}. This is a genetic implementation of a threshold device that, by acting deterministically, becomes activated only if the input signal $x$ crosses a particular limit (Fig.~1B). When the signal is stochastic, the response depends of course on the relationship between this threshold and the mean (and variance) of the underlying distribution $P(x)$ (considered for simplicity as a uniform distribution; Fig.~1B). A symmetric distribution centered on the threshold would thus originate equally likely the two output values ({\tt OFF/ON}) (i.e., one bit of information); while the same distribution centered above/below the threshold would produce biased responses (i.e., less than one bit of information).

However, the previous behavior can be affected by the extensive noise sources acting on biological circuits~\cite{mcadams1997,elowitz2002,lestas2010}. One could ask then to what extent the circuit is reliably representing the signal. To quantify how much information the response conveys about the input, we made use of MI~\cite{levchenko2014,bowsher2014} (Fig.~1A). We thus computed the response to a number of signals drawn from a fixed distribution and strength of intrinsic noise (black dots in subpanels of Fig.~1C), which allowed us to quantify the value of MI. This value changes with noise [main plot in Fig.~1C; the mean of $P(x)$ is above the threshold, red distribution in Fig.~1B]. For weak noise levels, the circuit works essentially as a deterministic switch, it is always $y=$ {\tt ON} as $x>$ threshold. For strong noise levels, the device cannot distinguish signal fluctuations, then its behavior is essentially random. In both cases, the information that the gene processes is limited (subpanels of Fig.~1C, red curves denote the averaged stimulus-response profiles). But MI presents a maximum for an intermediate noise level. In this regime, the circuit can express its two possible states due to noise (i.e., low values of $x$ can cross the threshold)~\cite{gammaitoni1995}, what precisely contributes to a better representation of the input signal (see also Fig.~S1); a characteristic behavior of noisy nonlinear systems known as stochastic resonance (SR)~\cite{gammaitoni1998}.

Moreover, SR disappears when the mean of $P(x)$ is close to the threshold, as stochasticity is now not required to reach the two possible states. In this case, noise always reduces information transfer (Fig.~1D, curves for $N=1$). Note here how MI does exhibit an upper limit of 1 bit when the mean of $P(x)$ exactly matches the threshold, and the circuit is noiseless. MI decreases with noise because signal values above/below the threshold originate in some cases stochastic crossings (e.g., $y=$ {\tt ON} when $x<$ threshold), and the information content in absence of noise is already high (note in contrast that, in the scenario of SR, MI was very low in absence of noise). Additionally, Figure~1D displays a situation in which a maximum in MI is nevertheless observed (curves for $N=2$). This is obtained by increasing the number of devices processing the same input, with $y$ representing in this case the sum of all individual outputs; a phenomenon called suprathreshold SR~\cite{stocks2000}. What is apparent here is that redundancy boosts information transfer, given a fixed noise level.

\subsection*{Genetic redundancy enhances information transfer} 

The addition of extra copies of the threshold device, i.e., genetic redundancy, appears then as a potential mechanism to increase the transmission of information in the presence of intrinsic noise. Consider, for instance, a situation in which two devices read in parallel the same input signal, assuming again two possible values of gene expression for each unit. The overall output {\it alphabet}~\cite{shannon1948} consists of three {\it letters}: \{{\tt 0} (both copies {\tt OFF}), {\tt 1} (one {\tt OFF} the other {\tt ON}), {\tt 2} (both {\tt ON})\}. The new alphabet is linked, of course, to the action of independent (intrinsic) noise sources acting on the two genes, which allows each device to produce an autonomous response (with noise-induced threshold crossings). The sum of individual responses would give, accordingly, a global output distribution $P(y)$ constituted by three peaks. The extended alphabet helps therefore to enlarge the capacity of the output to represent the input variability; in other words, it contributes to linearize the averaged stimulus-response profile (Fig.~2A, see also Fig.~S2).

Both the number of units and the type of nonlinearity influence the increment of information transfer. In Figure~2B, we introduced three different threshold devices~\cite{wall2004} to show how MI increases with redundancy. For each type, MI relative to the case of no redundancy (i.e., a single unit) was plotted. Specifically, we examined a simple regulated unit, a bistable expression system implemented through a positive feedback (the architecture that we discussed before), and an excitable device constituted by interlinked positive and negative feedbacks [implemented as the one linked to transient differentiation in {\it Bacillus subtilis}~\cite{suel2006}]. The output of all these devices is given by a continuous variable representing gene expression (note that the response was previously regarded as {\tt OFF/ON}). This allowed identifying discrepancies in terms of MI among different gene regulatory circuits. In particular, the largest amplification of information content corresponds to those devices whose actions ultimately rely on discontinuous transitions (i.e., the bistable and excitable systems). Out of these two systems, the excitable one presents comparatively larger amplification, although only observed for relatively large arrays. This is associated to the fact that, in this system, the response is entirely binary even in presence of noise: either the signal triggers a response or not (Fig.~S3). Moreover, the gain in information transfer is much lower for the simple regulated system. In this case, the stimulus-response profile is continuous (i.e., no discontinuous transition is produced) what entails that one unit already has the capacity to reach a relatively large output alphabet. The contribution of redundancy is therefore always much higher in analog-to-digital than in analog-to-analog signaling circuits (and provided they are noisy).

\subsection*{Input signal distribution shapes information transfer} 

The specific distribution of the signal impinging on the genetic circuits can encode specific environmental or genetic conditions~\cite{sharpe2001}, which can further modulate the enhancement of information transfer. We first analyzed the effect of the shape of $P(x)$. We considered three different signals acting on the array of threshold devices. A normal distribution contributes in higher extent to increase MI with genetic redundancy (Fig.~3A). For this distribution, the mass of $x$ values is closer to the threshold, existing more chances to subvert the deterministic decision of the device due to noise. We then analyzed the effect of the relationship between the threshold and the signal mean. When the mean of $P(x)$ is equal to the threshold, a higher increase of MI with genetic redundancy is observed (Fig.~3B). Arguably, if the mass of $x$ values is equally distributed above/below the threshold, there exists again more chances for noise-induced threshold crossings. Fine-tuning of the parameters characterizing $P(x)$ contributes thus to a better representation of the input signal by the global output response.

\subsection*{Extrinsic noise and cross-talk limit information transfer} 

The most important constraint for the gain in information associated to the previous redundant systems is the independence between the noise sources. When these are correlated, $P(y)$ becomes more sharply peaked around a small subset of possible responses (i.e., the output alphabet is more limited; Fig.~4A). This applies to biological circuits that, in addition to intrinsic noise, also integrate the effect of extrinsic fluctuations~\cite{elowitz2002,swain2002}. This type of noise affects all genetic devices in the same manner what eventually correlates individual outputs. Figure~4B shows how (relative) MI decreases with the strength of extrinsic noise in an array of five bistable units. Note however that this redundant architecture still exhibits, for different extrinsic noise levels, a larger MI with respect to the nonredundant case (inset of Fig.~4B).  

Despite the independence of the noise sources, the presence of cross-talks between devices can similarly lead to correlations in the individual gene responses. In a genetic context, one could imagine two independent transcription factors sharing recognition domains~\cite{masquilier1992}. One could also imagine a second unit recently emerged by duplication, and that no process of neofunctionalization yet occurred~\cite{hittinger2007}. Figure~4C indeed shows a decay in (relative) MI for a system of two units when cross-talk between them increases (simulations done without accounting for extrinsic noise). In this case, the activation of one unit drags the activation of the other, biasing again the output alphabet (inset of Fig.~4C). Of note, the decay profile in MI is qualitatively different in the two scenarios. Addition of extrinsic noise contributes to limit information transfer in a progressive manner since it increasingly coordinates responses. In the second scenario, outputs are correlated once a relatively specific cross-talk range is reached what is reflected in a more abrubt decay.

\subsection*{Heterogeneity also contributes to enhance information transfer} 

A complementary source of individuality in information processing could be linked to the heterogeneity within the collection of threshold devices. In the context of genetic circuits, this corresponds to the variability in promoter strengths, ribosome-binding sites, proteins half-lives, or protein-DNA binding affinities; all factors that in effect modify threshold values or output responses. Adjusting for each device the values of the biochemical parameters of the model can capture this variation~\cite{mayo2006}. We specifically explored the implication of threshold heterogeneity in the array of five bistable units. 

Notably, we observed again a resonance in information transfer, but this time as a function of the degree of heterogeneity (Fig.~5). While moderate levels of heterogeneity allows regulatory circuits to encode complementary aspects of the input signal, hence enhancing information transfer, larger levels of variation originates noise-induced threshold crossings over the whole input range, which is detrimental to represent $P(x)$ with $P(y)$ (note that these crossings occur in a narrower range when less variation is considered). Moreover, since both intrinsic noise and heterogeneity contribute to increase the transmission of information, we also explored to what extent these two sources of individuality work independently~\cite{hunsberger2014}. We found that intrinsic noise mitigates the increase in MI due to threshold variability (inset of Fig.~5). Intuitively, higher noise levels make indistinguishable those regulatory variations in terms of gene expression.

\section*{DISCUSSION}

Binary decisions implemented by means of threshold devices appear in many engineering and physical systems, and have been extensively studied in relation to the detection and transmission of signals. While noise was commonly considered harmful in many of these scenarios, some work alternatively identified circumstances in which its presence enhances performance~\cite{gammaitoni1998,mcdonnell2011}. In Biology, both the stochastic nature of biochemical reactions and the typical occurrence of thresholds --linked, for instance, to cell fate determination-- also anticipates the possibility of beneficial effects. This specifically applies to the case of gene regulatory circuits, in which molecular stochasticity acts in many cases as a core determinant of function~\cite{eldar2010}.

In this work we discussed in detail the benefits of intrinsic molecular noise when multiple threshold regulatory circuits process a common signal. This system exhibits a resonance phenomenon known as suprathreshold SR~\cite{stocks2000}. The effect establishes the benefit of the noise-induced uncoupling of the action of each unit. This advantage is manifested as well in a more linear relation between stimulus and response, a type of dose-response alignment that could be important in how precise extracellular conditions determine cell responses, and that was previously associated to negative feedbacks~\cite{richard2008}. Our functional analysis therefore reveals redundancies not only as a genetic architecture contributing to robustness~\cite{kafri2006,keane2014}, or to the adaptation to novel environments through the increase of gene expression levels~\cite{riehle2001,gresham2008}, but also as a mechanism increasing the capacity to transmit reliable information (Fig.~6). We suggest that this aspect could selectively contribute to the evolutionary maintenance of genetic redundancy. That multiple signaling pathways in {\it Saccharomyces cerevisiae} overlap supports this hypothesis~\cite{wageningen2010}.

The balance of intrinsic/extrinsic noise also plays an important part to condition the amount of information transferred (Fig.~6). Cells implementing regulatory circuits with few representative molecules or living in rich environments would shift this balance towards intrinsic noise~\cite{volfson2006}. Beyond this genetic/environmental tuning, cellular systems could avoid the loss of information, due to extrinsic noise, when the signal operates dynamically rather than statically~\cite{selimkhanov2014}. Note that here we considered a static operation. Our results further emphasize how heterogeneity and cross-talk among redundant copies play opposite roles in the maintenance of information content (Fig.~6). One could thus interpret the action of several parallel signaling pathways, each conveying approximately 1 bit of information, as heterogeneous copies of an effective threshold device what enhances information transmission, e.g., this was observed in pathways for the growth factor-mediated gene expression~\cite{uda2013}.

That a global response --the sum of individual responses, in this case-- implemented by parallel processing units could lead to better performance than that of the individual components was proposed in early models of computing, and can indeed be observed at different levels of biological organization: from genes (this work), to living cells~\cite{cheong2011}, to social organisms~\cite{conradt2005}. In addition, ideas on redundancy and heterogeneity when mounting unreliable components were already present in the initial development of fault-tolerant computation and communication~\cite{vonneumann1956,moore1956}, and also permeate to many biological scenarios. Our work substantiates the implications of these notions in cellular decision making by natural~\cite{wageningen2010} and synthetic~\cite{dueber2007} molecular circuits, and contributes to exemplify how the application of concepts from information theory could lead to a more precise and quantitative understanding of cellular systems.

\section*{THEORETICAL PROCEDURES}
\subsection*{Modeling noisy regulatory systems}
We considered a redundant system consisting of $N$ different transcriptional units, each of them activated by the input signal ($x$). The model for the $i$-th unit reads 
\begin{equation}
\frac{dy_i}{dt} = f(y_i,x) + q_i(y_i,x) \xi_i(t),
\label{model1}
\end{equation}
where expression ($y_i$) and time are appropriately rescaled to have a dimensionless model. To model different regulatory systems (simple, bistable or excitable), we modified the function $f$ (see the Supplement for details of functional forms and parameter values). $\xi_i$ is a stochastic process that has mean $0$ and is $\delta$-correlated. Noise amplitude is given by the square root of the sum of propensities.\\ 
\indent To account for extrinsic noise, we introduced a new stochastic process ($\xi_{ex}$), common to all units, in Eq. (\ref{model1}) as 
\begin{equation}
\frac{dy_i}{dt} = f(y_i,x) + q_i(y_i,x) \xi_i(t) + q_{ex} \xi_{ex}(t).
\label{model2}
\end{equation}
The correlation time of extrinsic noise is of the order of the cell cycle (the mean is also $0$). For simplicity, we here supposed a system implemented with short-lived proteins, so we can assume that $\xi_{ex}$ is constant within the time window that the system needs to reach its steady state upon receiving the perturbation $x$.\\
\indent To account for certain cross-talk between the different units of the system, we followed a perturbative approach to obtain
\begin{equation}
\frac{dy_i}{dt} = f(y_i,x) + q_i(y_i,x) \xi_i(t) + \varepsilon \sum_{j=1, j \neq i}^N y_j,
\label{model3}
\end{equation}
where $\varepsilon$ quantifies the degree of cross-talk. For simplicity, we assumed $q_i$ not to be dependent on $y_j$ for $j \neq i$. \\
\indent To account for heterogeneity, we included variations in the threshold values of the different units of the system. This was modeled by introducing a Gaussian random number $\omega$ of mean $1$, being its standard deviation the degree of heterogeneity [$f(y_i,x,\omega_i)$ for the $i$-th unit, where $\omega_i$ is a realization; see details in the Supplement]. When accounting for cross-talk or heterogeneity, only the intrinsic noise was considered.

\subsection*{Input and output variables}
Here, we contemplated that the regulatory system is initially in a steady state in which there is no input signal (i.e., $x=0$ for $t<0$). We then considered that $x$ becomes activated (at $t=0$); as a step function in the case of the simple and bistable systems, or as a pulse function (for one unit of normalized time) in the case of the excitable system. The value of $x$ for $t>0$ was modeled as a random number following a given probability distribution. This models an input signal whose value can fluctuate according to upstream processes, environmental changes or molecular noise. We also regarded that the array of genes is able to perceive this signal to change the individual expression levels ($y_i$) accordingly. The output was calculated at steady state. We assumed that the signal fluctuations occur at a frequency that allows the genetic circuit to respond against the current signal value. \\
\indent The change in gene expression due to signaling was defined by $\Delta y_i = y_i(x) - y_i(x=0)$. The total differential gene expression of the redundant system can be written as $\Delta y = \sum_{i=1}^N \Delta y_i$. In case of the excitable system, because the response is transient, we considered a Boolean function operating on $y_i$, setting $1$ if the unit was excited or $0$ if not. In the first section of the paper (Fig.~1), the gene expression level ($y_i$) was treated as a Boolean variable ({\tt OFF/ON}). In the subsequent sections (Figs.~2 -- 5), it was treated as a continuous variable. \\
\indent In addition, we studied different distributions of $x$. We mainly included a uniform distribution covering two orders of magnitude. In Figs.~1 and 3B, we analyzed the effect of the mean of the distribution, and the values were $0.001$ (equal to the threshold value), $0.005$ and $0.01$. In Fig.~2, the mean was fixed to the threshold value, i.e., $0.001$ in the case of the bistable system, $1$ in the simple regulated unit, and $0.9$ in the excitable system. In Figs.~4 and 5, concerning to the bistable system, the mean of the distribution of $x$ was $0.005$. We additionally considered different forms of the distribution. In Fig.~3A, we analyzed the effect a normal or beta distributions in log scale, with the mean equal to the threshold value.

\subsection*{Quantification of information transfer}
We used mutual information (${\cal I}$) as a quantitative metric to describe how the global output response of a single cell is sensitive to different concentrations of the input signal. This extends the quantification by the averaged stimulus-response profile. To calculate ${\cal I}$, we performed $10^4$ realizations of the pair ($x$, $y$) and then we solved numerically the following integral
\begin{eqnarray}
{\cal I} &=& -\int_{-\infty}^{+\infty} P_{\Delta y}(s) \log_2 P_{\Delta y}(s) ds + \int_{-\infty}^{+\infty}P_{\log x}(r)  \nonumber \\ &\times&  \int_{-\infty}^{+\infty} P_{\Delta y | \log x}(s) \log_2 P_{\Delta y | \log x}(s) ds dr,
\label{mi}
\end{eqnarray}
where we considered $\log x$ as input and $\Delta y$ as output variables. By using the Fokker-Planck equation, we calculated the probability that a unit has a given gene expression level (see more details in the Supplement). 

\section*{Acknowledgments}
We thank M.A. Fares and A. Couce for a critical reading of the manuscript. This work was supported by grants (to GR and JFP) from the Spanish Ministry of Economy and Competitiveness. 

\section*{Author Contributions}
G.R. and J.F.P. performed research and wrote the manuscript.

\section*{Competing Financial Interests}
The authors declare no competing financial interests.

\bibliography{redunoise_refs}

\newpage

\begin{figure*}
\label{Fig1}
\centerline{\includegraphics[width=0.76\textwidth]{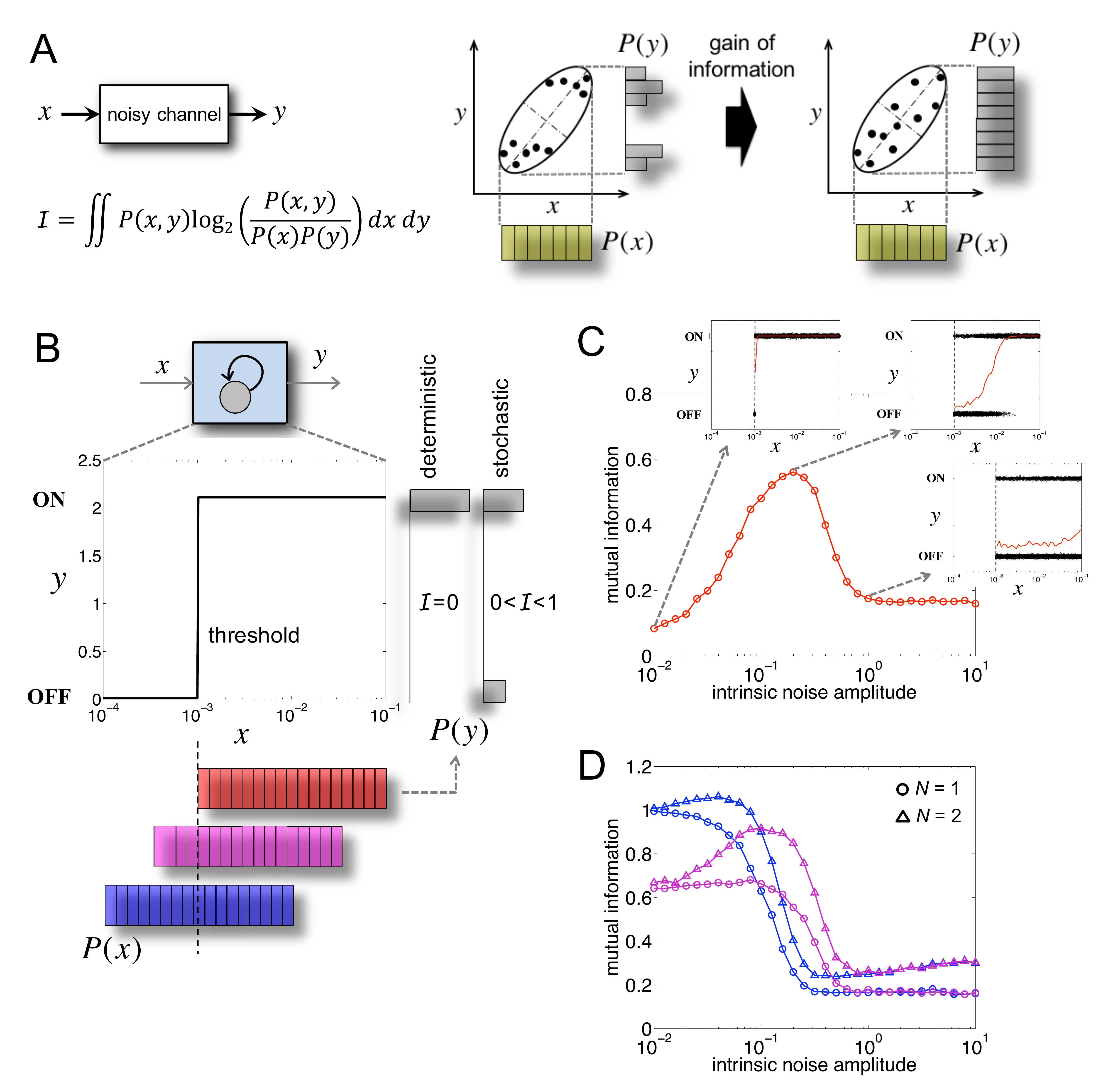}}
\caption{\small{{\bf Intrinsic noise can increase or decrease information transfer in threshold genetic systems.} (A) A noisy channel is characterized by the mutual information (MI) ${\cal I}$ of the output ($y$) given an input ($x$). MI quantifies the dependence between input and output distributions, $P(x)$ and $P(y)$ respectively. This could be estimated by a correlation coefficient, but this measure cannot discriminate some associations better captured by MI. In the cartoon we show two cases with the same correlation (whose value we represented here by the eccentricity of the ellipses) but different MI. (B) The channel can describe a gene autoactivating its own expression ($y$) in a bistable {\tt OFF/ON} manner what represents a simple example of threshold regulatory circuit. Information transfer depends on the relationship between $x$ and the threshold value of activation ($x$ and $y$ are presented in arbitrary units). Three instances of $P(x)$ are shown (uniform distributions with different means; the blue one corresponds to a mean equal to the threshold value). When the signal is always beyond the threshold (red distribution) the circuit exhibit a nonzero MI only when it works stochastically (note the two different output distributions). Here we considered a binary response ({\tt OFF} if $y<1$, {\tt ON} otherwise). (C) Resonance in MI as a function of the strength of intrinsic noise (see Theoretical Procedures) for the red $P(x)$ in (B). Each subplot displays the responses of the device to $10^4$ signal values drawn from the described distribution (black dots), and the corresponding averaged stimulus-response profile (red curve), for three explicit noise levels. The maximum in MI occurs when the averaged stimulus-response profile is more linear (Fig.~S1). (D) Other signal distributions, in which intrinsic noise always reduces MI, can nevertheless exhibit a resonance when the combined response of several units is considered (we show here the case of duplicated threshold devices; $N=2$). Colors correspond to those distributions shown in (B). 
}} 
\end{figure*}

\begin{figure}
\label{Fig2}
\centerline{\includegraphics[width=0.7\textwidth]{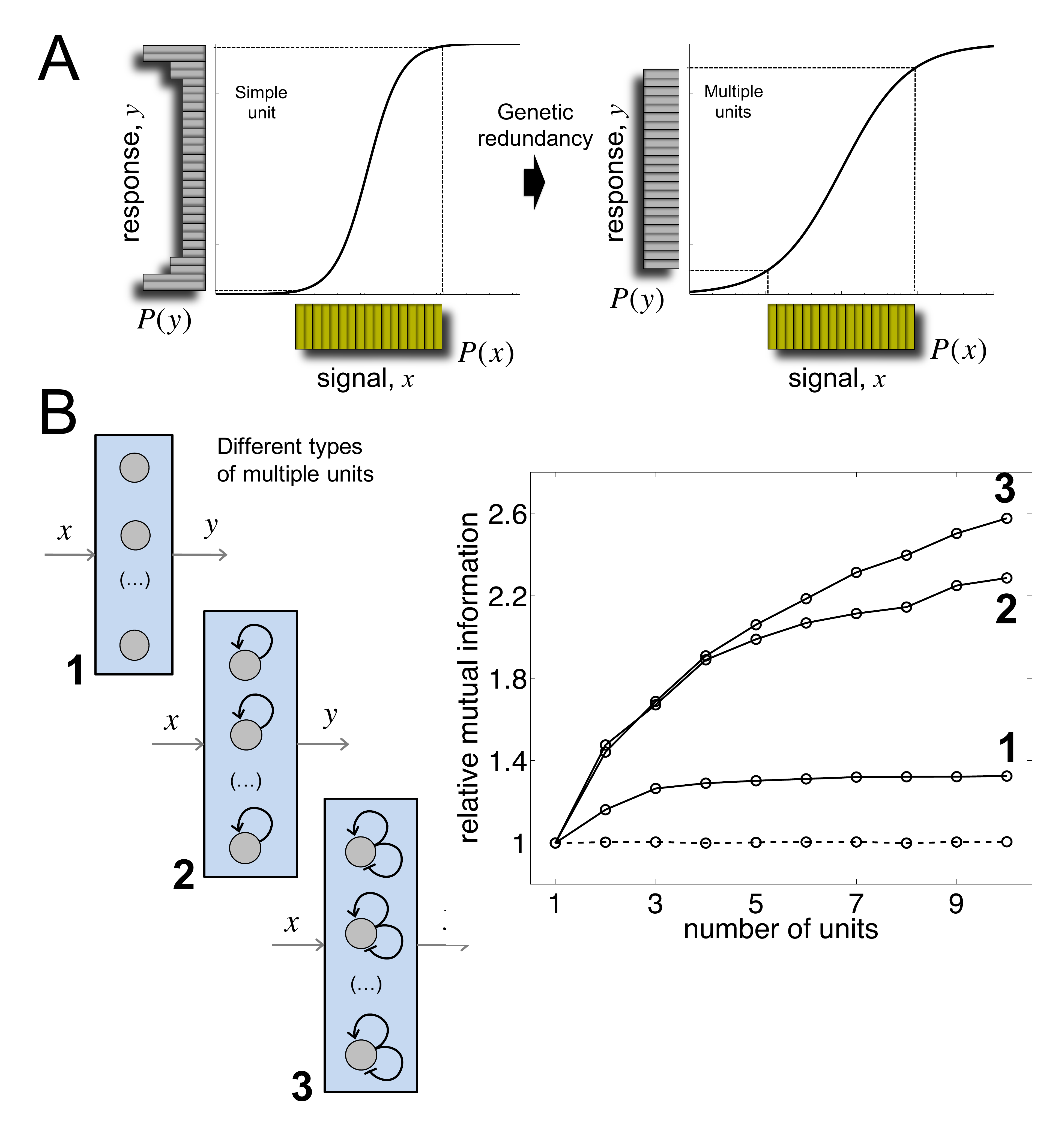}}
\caption{\small{{\bf Genetic redundancy amplifies information transfer in threshold genetic systems.} (A) Input/output distributions depicting information transfer. The input distribution (in yellow) is assumed to be uniform. Output distributions (in gray) illustrate the processing of the signal $x$, either through a single copy of the threshold device (left) or an array of multiple redundant copies (right). In the latter case, each unit of the array receives the same signal and the output $y$ is the sum of all the individual responses. Redundancy effectively enlarges the alphabet of the response. This is reflected in the output distribution, and also in the linearization of the averaged stimulus-response profile (black curve). (B) (Left) Array of $N$ threshold devices whose constituent units correspond to (1) a simple regulated unit, (2) a bistable circuit implemented with a positive feedback, and (3) an excitable circuit constituted by two interlinked positive and negative feedback loops. (Right) Dependence of mutual information (MI) with the number of units ($N$) for each of these systems. MI relative to the case $N=1$. A uniform signal distribution with mean equal to the threshold value was considered. In the case of noiseless units, MI does not increase with extra copies (independently of the type of unit; dashed line). See the Supplement for details of the model of each circuit.
}}
\end{figure}

\begin{figure}
\label{Fig3}
\centerline{\includegraphics[width=0.9\textwidth]{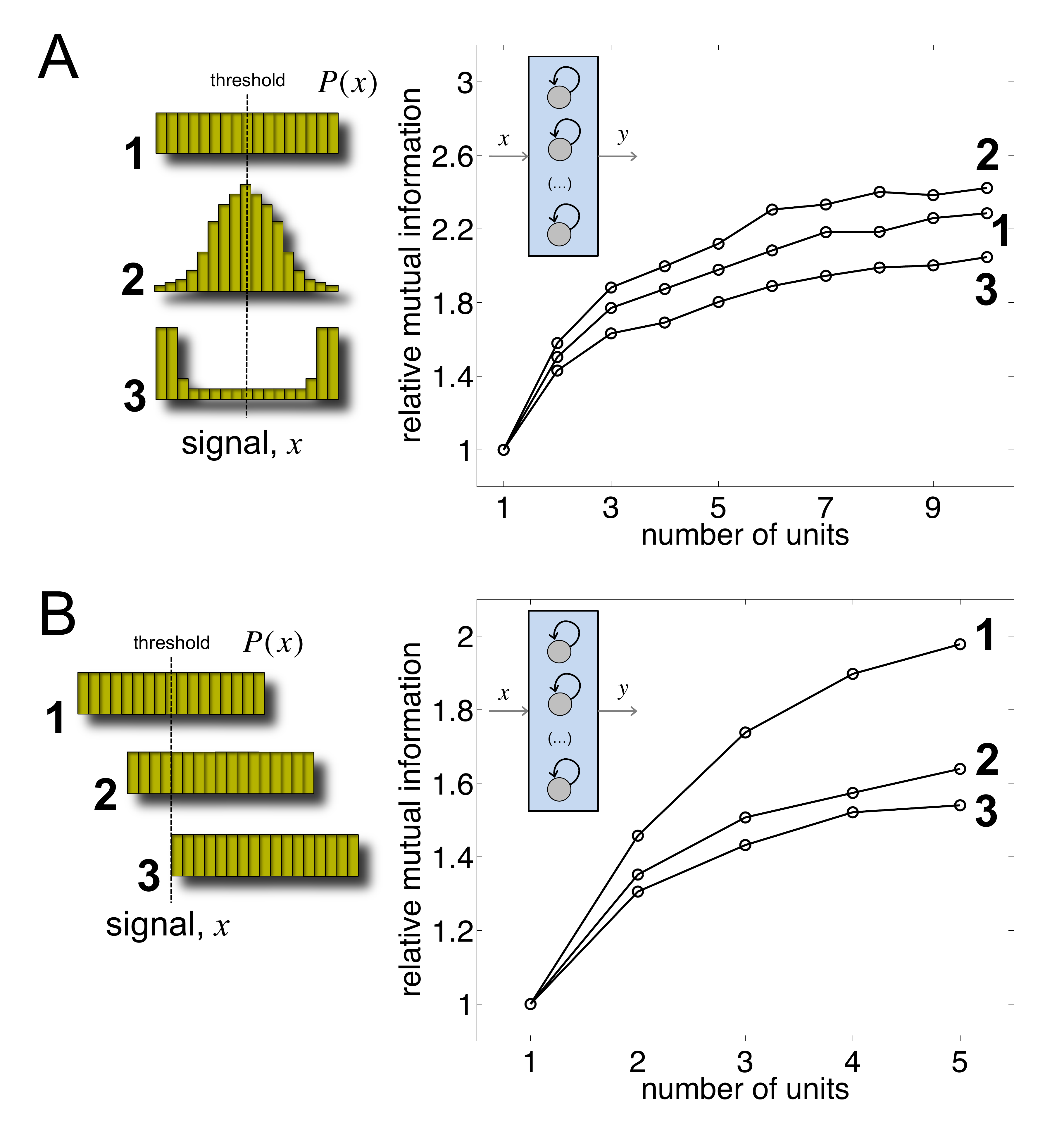}}
\caption{\small{{\bf The distribution of the signal modulates the increase of information transfer due to genetic redundancy.} (A) Effect of the form of the distribution on MI: (1) uniform (covering two orders of magnitude), (2) lognormal (with standard deviation equal to 2/3), and (3) beta in log scale (with the two shape parameters equal to 1/3). In all cases, the mean of the distribution is equal to the threshold value. (B) Effect of the mean of the distribution (here uniform) on MI: (1) equal to the threshold value, (2) and (3) deviated from the threshold value. We considered as threshold device a bistable unit implemented with a positive feedback in all plots (see Theoretical Procedures). 
}}
\end{figure}

\begin{figure}
\label{Fig4}
\centerline{\includegraphics[width=0.5\textwidth]{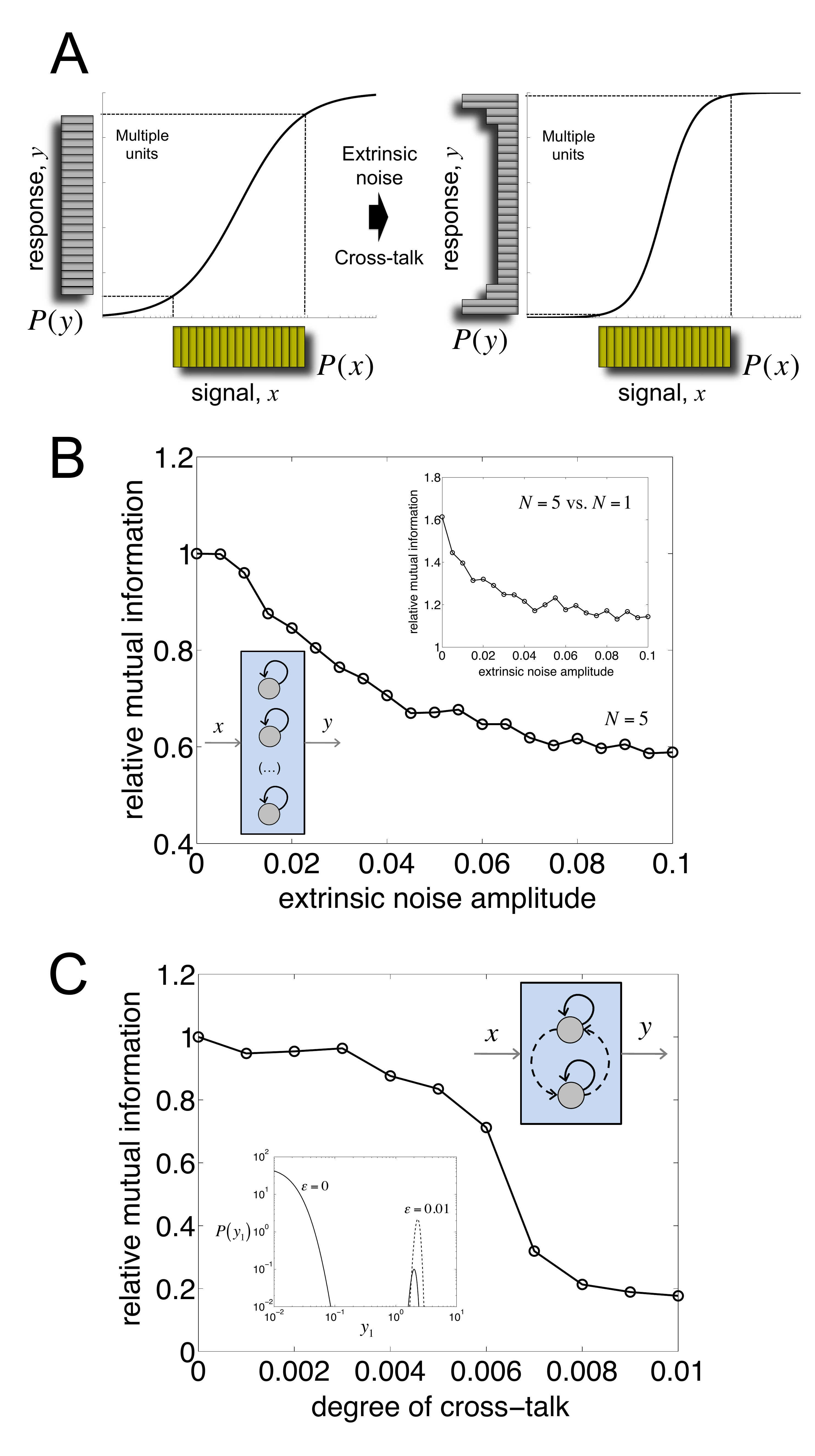}}
\caption{\small{{\bf Extrinsic noise and cross-talk among redundant copies limit information transfer.} (A) Input/output distributions depicting information transfer. In this case, correlation among individual gene responses due to extrinsic noise or cross-talk reduces the response alphabet, and generates a less linear averaged stimulus-response profile (black curve, see Fig.~2A for comparison). (B) Dependence of mutual information (MI) with the strength of extrinsic noise (see Theoretical Procedures). Relative MI is with respect to absence of extrinsic noise. For this plot, we considered a system of $N=5$ bistable units implemented with positive feedback. The inset shows a direct comparison between $N=1$ and $N=5$, emphasizing that MI increases with $N$. (C) Dependence of MI with the degree of cross-talk for the same regulatory system, but now constituted by $N=2$ units. Relative MI is with respect to the situation without cross-talk. The inset presents the marginal probability distribution of gene expression of one unit ($y_1$) in the absence and presence of cross-talk (parameterized by $\varepsilon$=0 and $\varepsilon$=0.01, respectively; see Theoretical Procedures) for the mean value of the input signal ($x$). Note that when the units are coupled, gene expression becomes unimodal (dashed curve).  
}}
\end{figure}

\begin{figure}
\label{Fig5}
\centerline{\includegraphics[width=0.9\textwidth]{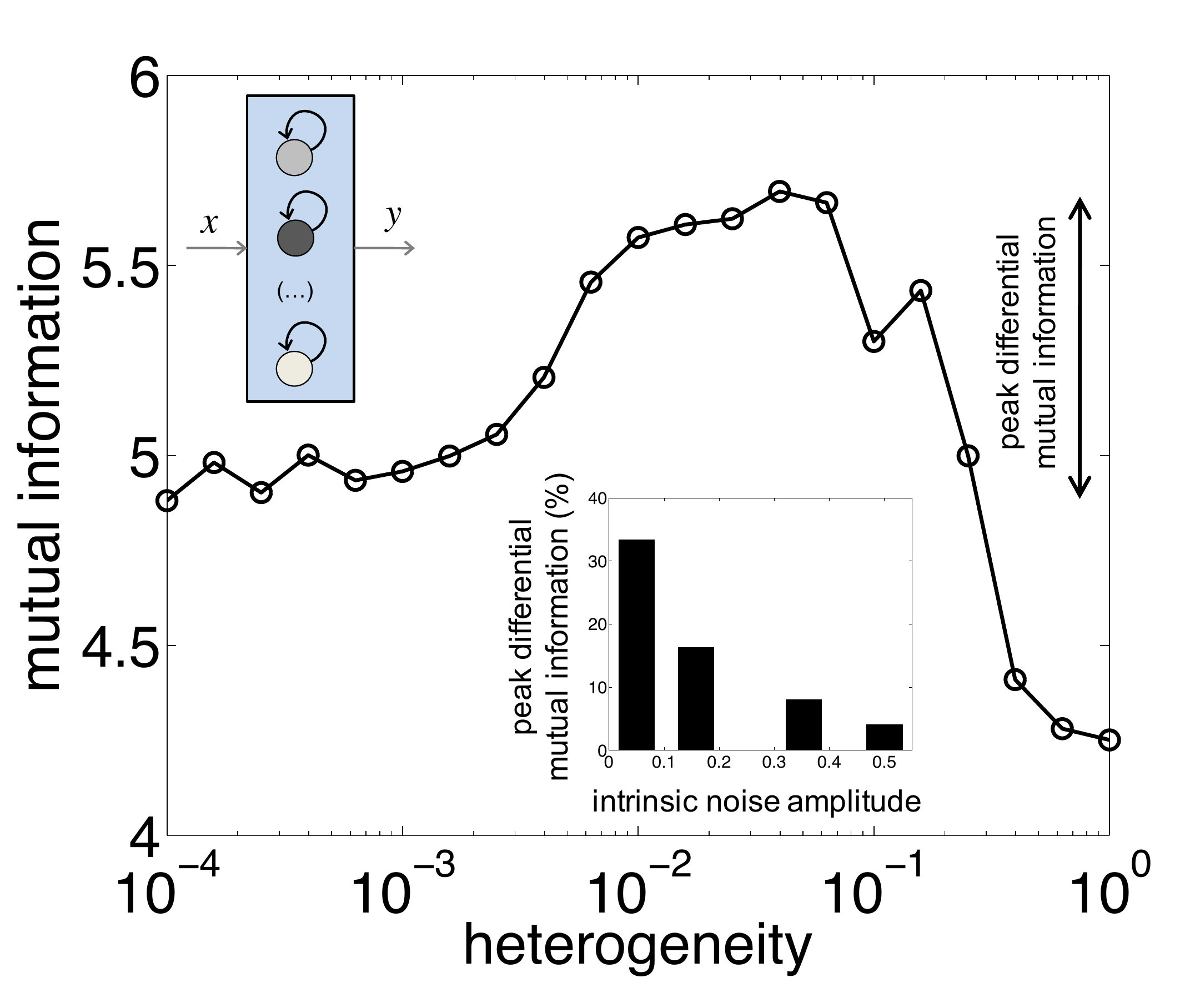}}
\caption{\small{{\bf Genetic heterogeneity among redundant copies leading to functional variability improves information transfer}. Variation in the biochemical features of the constituent threshold devices (here bistable units, $N = 5$; see Theoretical Procedures) leads to a maximum in mutual information (MI). The inset indicates the peak differential MI (i.e., the difference between the largest value of MI with heterogeneity and the value of MI without it) for varying noise levels. This reveals how a situation of stronger intrinsic noise contributes to reduce the improving effect on MI of heterogeneous units (the main plot corresponds to an intrinsic noise amplitude equal to 0.16). 
}}
\end{figure}

\begin{figure}
\label{Fig6}
\centerline{\includegraphics[width=1\textwidth]{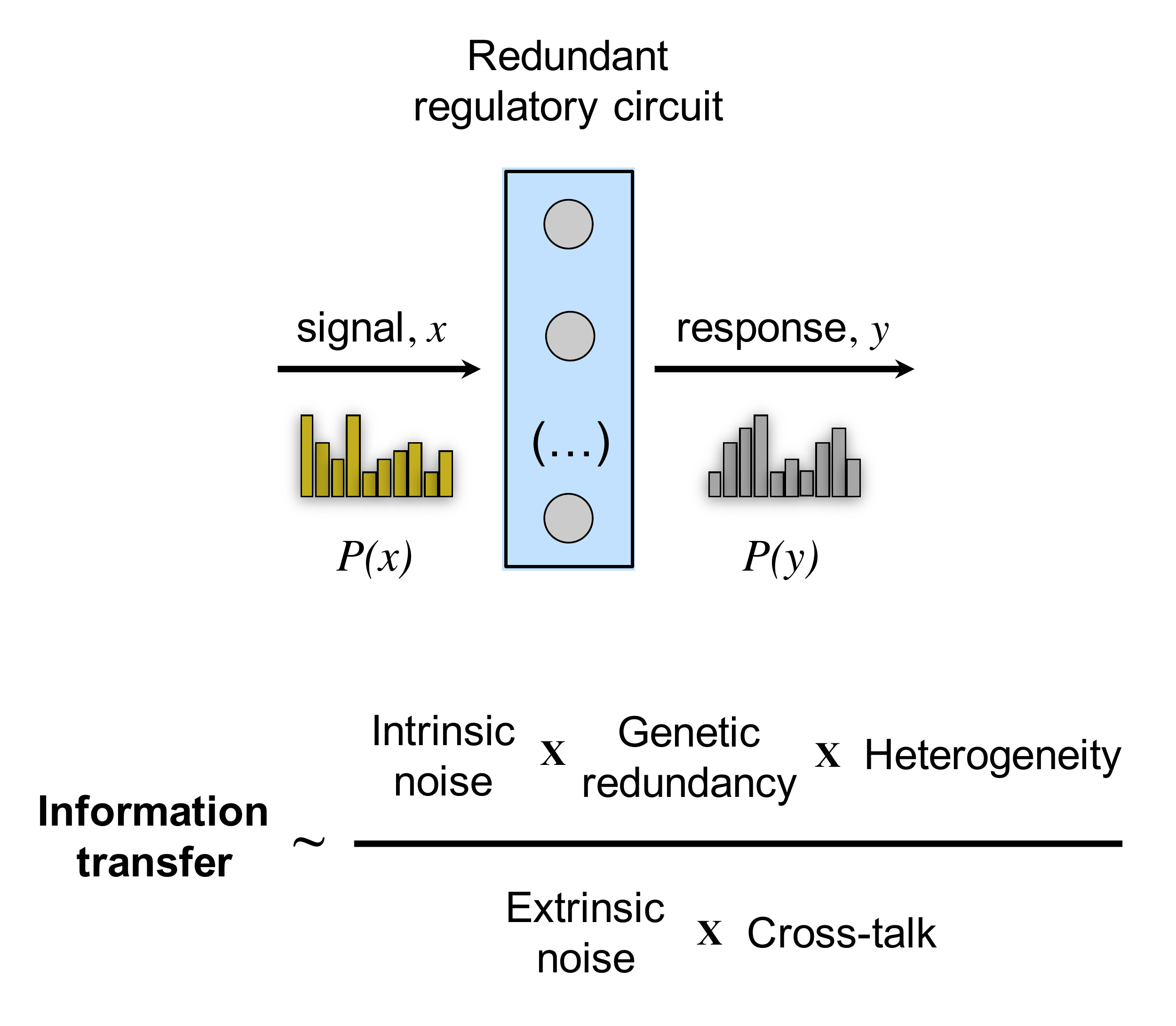}}
\caption{\small{{\bf Model of information transfer in gene regulatory circuits.} Intrinsic noise, genetic redundancy, and heterogeneity increase the transmission of information (by strengthening the capacity of the global output to represent the input variability), whilst extrinsic noise and cross-talk among redundant units become limiting factors (by correlating the individual outputs of the units).
}}
\end{figure}

\end{document}